\documentclass[fleqn,twoside]{article}
\usepackage[headings]{espcrc2}

\readRCS
$Id: espcrc2.tex,v 1.2 2004/02/24 11:22:11 spepping Exp $
\usepackage{epsfig,balance,flushend}
\usepackage[figuresright]{rotating}

\newcommand{\AmS}{{\protect\the\textfont2
  A\kern-.1667em\lower.5ex\hbox{M}\kern-.125emS}}

\title{\textbf{Cache Discovery Over a Multihop Wireless Ad Hoc Network}}
\author{Preetha Theresa Joy\address{Department of Computer Science, 
Cochin University of Science and Technology, Kochi, \\~Kerala, India.
Contact: preetha@mec.ac.in \\},
K Poulose Jacob\addressmark
}

\runtitle{Cache Discovery Over a Multihop Wireless Ad Hoc Networks}
\runauthor{Preetha Theresa Joy amd K Poulose Jacob}

\begin{document}
\begin{abstract}
Multihop ad hoc wireless networks consist of mobile nodes that communicate with each other without any fixed infrastructure. The nodes in these networks are power constrained, since they operate in limited battery energy.  Cooperative caching is an attractive solution for reducing network traffic and bandwidth demands in mobile ad hoc networks. Deploying caches in mobile nodes can reduce the overall traffic considerably. Cache hits eliminate the need to contact the data source frequently, which avoids additional network overhead. In this paper we propose a cache discovery policy for cooperative caching, which reduces the power usage, caching overhead and delay. This is done by power control and transmission range adjustment. A cache discovery process based on position coordinates of neighboring nodes is developed for this. The simulation results gives a promising result based on the metrics of studies.  
\end{abstract}

\maketitle

\section{\uppercase{Introduction}}
In mobile ad hoc networks (MANET)s, no base stations exist and each
mobile client act as router and packet forwarder. Networks can be
formed and fragmented on the fly without the intervention of a system
administrator or the presence of fixed network devices. Ad hoc
networks have multiple applications in the areas where wired
infrastructure may be unavailable such as battle fields and rescue
areas. In these types of networks new hosts can appear and old ones
can disappear at any time. The topology of this network is very
fragile; it can change at any moment and disconnections are frequent
due to mobility or activity status changes. Mobile hosts are powered
by battery while they are on move. 
\vskip 2mm
Thus, to ensure good continuity of
system operations over time, several approaches are taken to enhance
the battery life. One such approach is to design power aware
transactions, which make efficient use of the overall energy resources
of the network. Conserving power prolongs the life time of a node and
also the life time of the network as a whole. The transmit power of a
node can be adjusted to achieve maximum possible power savings. From a
data-management point of view, these restrictions introduce several
issues that need to be addressed. The mobile clients may suffer from
long access delay or access failure, when the nodes holding the data
items are far away or unreachable. Data transfers must be reduced and
mechanisms must be deployed to address the frequent disconnections and
low bandwidth constraints.
\vskip 2mm
Data caching is recognized as a feasible approach to improve the
performance in many traditional systems. Caching is the process of
pre-fetching the needed data and storing it closer to the source. In
mobile ad hoc networks, caching can improve mobile client perception
in three ways. First, retrieving data from a remote data center
involves wireless media network transfers and there is a chance of
data loss due to the wireless link characteristics. Second, when the
data is served locally, the network latency is reduced. The data
server processes the data request only when there is a local and
cooperative cache miss. By doing this the server load is balanced,
which consequently reduces the latency in serving client
request. Thirdly, frequent disconnections which occur in ad hoc
networks can be hidden from users, making the network more
reliable. Data caching in ad hoc networks are mainly proposed as
cooperative caching \cite{r6}.
 \subsection{Cooperative Caching}
Cooperative caching is the sharing and coordination of cache state
among multiple clients and has been recognized as an important
technique to reduce data traffic and to alleviate network bottlenecks
in ad hoc networks [1-22]. The mobile
clients can communicate among themselves to share information rather
than relying on the server. In MANET, it is likely that multiple
clients in the same region will try to access the same service
concurrently. So caching such services would be beneficial. For
example, an ad hoc network formed by the participants in a conference
may have common research interests. They may download similar
documents from the server. If anyone of the mobile client downloads a
document from the server, all other neighbors who are interested in
this document can retrieve it from the mobile client without
contacting the server. 
\vskip 2mm
As a result, there is a growing interest in
research in this field that has resulted in caching policies suited
for the characteristics of MANETS. Cooperative caching aims to reduce
the redundant data transfer using a mechanism which enables the local
cache of different mobile clients to be shared in a cooperative
manner. In cooperative caching the mobile clients are configured to
request the data object from its local set of data items, if not found
it queries its neighboring nodes. When there is a cache miss in the
neighboring nodes queried, the data item is retrieved directly from
the server and this procedure continues recursively.
\vskip 2mm
Cooperative caches have different functions: Cache discovery,
placement and replacement, consistency maintenance and data
dissemination. Discovery refers to how a mobile node locates the
cached data. Placement is the processes of selecting appropriate nodes
as cache locations and replacement is the strategy used for evicting
data when the cache is full. Consistency maintenance is maintaining
consistency among the source data and cached copies.
\vskip 2mm
Efficiency of a cache discovery protocol may be measured using
different metrics. Some of them include communication overhead, energy
consumption, delivery success rate and response delay. Towards the
goal of improving the performance in cooperative cache, we propose a
cache discovery scheme in which each node is able to dynamically
adjust its transmission range to reach its neighbouring nodes, thus
saving power whenever possible. 
\vskip 2mm
The remainder of the paper is structured as follows. Section 2 gives
an overview of different cache discovery process available, Section 3
reviews related works in cache discovery, Section 4 presents the
energy model, Section 5 explains the system design of the proposed
cooperative caching scheme, Section 6 describes simulation model and
implementation details, Section 7 describes the  experimental results
and Section 8 concludes the paper.
\section{\uppercase{Overview of Cache Discovery Approaches}}
Two approaches used for information discovery in cooperative cache
include broadcast-based approach and cluster-based approach. Although
first method is the simpler version, it relays on flooding to
broadcast the data request. Flooding increases network contention and
overhead when the network density is high. In contrast, cluster-based
approach has been bestowed with the features of reduced overhead by
having a coordinator node which manages the discovery process.
\vskip 2mm
 The neighbouring node which possesses the data can be easily found out by
checking the lookup table maintained by the cluster head.  The
disadvantage of this approach is that group maintenance is difficult
due to the mobility of nodes. The control node may get disconnected
which causes excessive overhead \cite{r4}. The number of entries in
the look up table increases when the network density is high. To
maintain the correct status of the network, these tables must be
frequently updated. This involves information exchange between the
nodes which in turn increases the traffic overload in a dense network.
\vskip 2mm
To circumvent these drawbacks we designed an energy-efficient cache
discovery protocol that minimizes the energy consumption for cache
discovery and maximizes the life time of nodes. 
\section{\uppercase{Related Work}}
Caching data in mobile nodes is an effective technique to improve
performance in a mobile environment. Recently, several schemes for
cooperative caching in mobile ad hoc networks have been presented in
the literature. The algorithms proposed in \cite{r7}, \cite{r8},
\cite{r15}, focuses more on cache placement and discovery
while\cite{r6}, \cite{r9}, concentrates more on cache management
protocols. Cache management includes cache admission and
replacement. The work \cite{r2}, \cite{r4}, \cite{r5}, \cite{r11},
\cite{r10} considers both aspects in cooperative caching. Below, we
describe some representative cache discovery schemes for cooperative
caching. 
\vskip 2mm
Lim {\it et al.}, \cite{r5} proposed an aggregate caching scheme to increase
the data availability in Internet based mobile network. They used a
broadcast based information search algorithm called simple search to
locate the required data item. Whenever a mobile node needs data, the
request is broadcasted to its adjacent nodes. Upon receiving the
broadcast request, the adjacent nodes reply to the request if it has
already cached the data, otherwise the request is forwarded to its
neighbours until it is acknowledged by an access point or some other
nodes which have the requested data. Flooding is the technique used
for broadcasting. This algorithm sets a hop limit for the request
packet to reduce the traffic in the network.
\vskip 2mm
 A caching technique which
uses cluster-based approach can be seen in \cite{r9}, in which a
coordinator node maintains the cluster cache state information of
different nodes within its cluster domain. If there is a local cache
miss, the coordinator node will determine whether the data item is
cached in other clients within its home cluster. Another approach for
data discovery other than the mentioned schemes can be seen in
\cite{r4} and \cite{r10}. Group caching \cite{r4} uses a table based
approach for cache discovery and data dissemination. Each node
maintains two tables, a group table and self table to maintain the
status of neighboring caching nodes.  
\vskip 2mm
 For cache discovery, the local
cache table is searched first and if data is not found, the group
table is searched to find the location of the cached data. The group
table contains cached data id and the node id which contains the data.
LRU is the replacement policy used. The drawback of this approach is
that the message overhead increases when the node density is
high. Also individual nodes have to process these messages which
increase the computational overhead. 
\vskip 2mm
In \cite{r10}, a cache discovery technique based on adaptive flooding
broadcast is used for searching data in the network. According to this
scheme a mobile node uses three schemes; adaptive flooding,
profile-based resolution and road side resolution. In adaptive
flooding, a node uses constrained flooding to search for items within
the neighbourhood. In profile-based resolution, a node uses the past
history of received requests. In road side resolution, forwarding
nodes caching the requested item, reply to the requests instead of
forwarding them to the remote data source. COCAS \cite{r15} is a
distributed caching scheme designed for MANETs to find the requested
data from cached nodes. 
\vskip 2mm
The submitted queries are cached in some
special nodes called query directories (QD) and these queries are used
as an index to find the previously cached data. Whenever a data item
is retrieved, cache nodes (CN) cache the data and the nearest QD to
the cache node will cache the query along with the address of the CNs
containing the corresponding data. The assignment of QDs and CNs are
done by a service manager. The limitations of this scheme include,
broadcasting of requests for searching QDs, and the single point of
failure of the service manager. 
\section{\uppercase{Energy Model}}
There are three major causes of energy consumption in a node
\cite{r21}.  Energy consumed while transmitting a message, energy
consumed while receiving a message and when a node is `on' and is not
actively receiving. In the transmitting mode, energy is consumed in
two ways: For message processing and transmission. In the receiving
mode, energy is consumed only for processing. Finally, in the `on'
mode energy consumption is for processing, but it is quite low
compared to the transmitting and receiving modes. Thus, to reduce
energy depletion in a node the number of transmissions should be
reduced.
\vskip 2mm
While transmitting a message, a node spends part of its energy and there are a few energy models used to compute this consumption. In the most commonly used one, the measurement of energy consumption when transmitting a unit message depends on the range of the emitter $n$:
\begin{equation}\label{eq1}
E(n)= r(n)^\alpha
\end{equation}
where $\alpha$ is the real constant greater or equal than 2 and $r(n)$
is the range of the transmitting node. Our objective in the proposed
cache discovery protocol is to reduce the number of messages involved
in cache discovery and uses a range adjustment mechanism to save
power.
\section{\uppercase{System  Design}}
\subsection{Network Model}
A mobile ad hoc network is abstracted as a graph $G (V, E)$, where $V$
is the set of nodes and $E\subseteq V^2$   is the set of links which
gives the available communication. An edge $(u, v)$ belongs to $E$
means that there is direct communication between two nodes $u$ and
$v$. The elements of $E$ depend on the position and the communication
range of nodes. All links in the graph are bidirectional i.e., if $u$
is in the transmission range of $v$, $v$ is also in the transmission
range of $u$. The maximum communication range is assumed to be same
for all nodes and is represented as $R$, which is given by the
Euclidean distance $d (u, v)$ between nodes $u$ and $v$. The set of
neighborhood nodes in the range $R$ are represented as $N_R (U)$ and
the set of neighborhood nodes in the range $R_i$, is given as
$N_{{\rm Ri}} (U)$.
\subsection{System Model} 
We assume a mobile ad hoc network with a set of nodes which are able
to communicate with each other. The transmission radius $R$ determines
the maximum communication range of each node and is equal for all
nodes in the network. Two nodes in the network are neighbors if the
Euclidean distance between their coordinates in the network is at most
$R$. The Euclidean distance between the nodes are estimated based on
the relative position of nodes. We assume that each node knows its
current location precisely with the availability of Global Positioning
System (GPS). For power adjustment we make use of path loss model. In
this model, the path loss depends on the height $s$ of the antennas as
well as the transmitter-receiver separation \cite{r22}.
\vskip 2mm
Initially, to find the neighbor node set in the transmission range $R$
for node $A$ $N_R (A)$, a short neighbor request control message is
disseminated in to the network. The request control message contains
the following fields: \textit{the source id}, \textit{current
  location} and \textit{a request id}. The request id is used to
identify the neighbor request control message. When a node receives
the request control message, it sends back a reply control message
which includes the \textit{node id} and \textit{current location
  coordinates}. Upon receiving the location coordinates of neighboring
nodes, the source node measures the distance $D_{ij}$ between the
source node $n_i$ and neighbor $n_j$ using the Euclidean distance
formula.
\vskip 2mm
When a node is farther away from the source the Euclidean distance
will be large. Each node will arrange the neighboring node list in the
ascending order of their distance to determine the order of
neighboring nodes to receive the data request. To reduce the number of
messages, the neighboring node set is divided into different zones,
based on the transmission range. To maintain the neighbor node set
accurately, each node periodically sends a request control message to
its neighbors. 
\vskip 2mm
Each node maintains a list which stores the cached data item. The list
contains the following fields: cached data id, cached data item, ttl,
time difference between the current access and previous access. This
table is updated when a new data is placed in to the cache .The cache
space for each node is limited and when it is full, a replacement
strategy evicts the unwanted data. The contents of the local cache are
shared by its neighboring nodes. 
\vskip 2mm
The data server is assumed to be a fixed location. The data server
maintains a set of data items uniquely identified by means of data
item id $D_i$   for $1\leq i\leq n$ where `$n$' is the size of the
data base. The size of each data item varies from $S_{{\rm min}}$ to
$S_{{\rm max}}$. Each node has a local cache, with certain data
items. Each mobile nodes is identified by a distinct $<$Host id,
Name$>$ for $1\leq i \leq N$, where $N$ is the density of the
network. Nodes in the network retrieve data items either from the
local cache or from the neighboring cache if there is a local
miss. When a node fails to find data in neighboring nodes, data is
retrieved from the data center. When a node receives a fresh data
directly from the server, it caches a copy of it in the local cache
and becomes a provider for that cached content for the neighboring
nodes.
\vskip 2mm
When a node wants to access data, it checks in its own local cache. If
the requested data is not cached, the node checks whether the data is
present in the neighboring nodes. If we are not able to find the data
from the neighbor list the request is given to the data server.  
\subsection{Proposed  Cache Discovery Algorithm}
The cache discovery protocol we propose is based on minimizing the
power or energy per bit required to transmit a packet from source to
destination. The goal of this scheme is to reduce the average number
of messages among the cooperative caches while maintaining high cache
hit ratio. The link coast for the transmission can be defined for two
cases, (a) when the transmit power is fixed and (b) when the transmit
power is varied dynamically as a function of the distance between
transmitter and the intended receiver. For the first case, the power
needed to transmit and receive a message depends on the message
size. For the latter case, the power consumed $P(d)$  by a node in
transmitting a request for a distance $d$ is given by 
\begin{equation}\label{new}
P(d) =d^\alpha +c
\end{equation}
for some constants $\alpha$ and $c$. If we ignore the constant we can
see that the power consumption is directly proportional to
distance. If the nodes can adjust their transmission power for each
node based on distance, the power consumption can be reduced
considerably, which leads to increased battery life. So by making use
of the position coordinates we can transmit packets with minimum
required transmit energy. The basic requirement of this scheme is that
each node should know its relative position.
\vskip 2mm
In the proposed algorithm, the decision to forward the data request is
based solely on the location of itself and its neighboring nodes. We
divide the maximum transmission range of a node in to different zones
with transmission radius $R/2$, $R/4$ and $R/6$ and find the nodes
present in this transmission radius excluding the one already present
in the lower range. This can be found from the neighbor node list
formed when a node is active.
\vskip 2mm
When the requested data is not found in the local cache the request is
forwarded to the nodes in the lowest transmission radius zone. After
sending the request the node waits for the reply. If the node doesn't
receive a positive reply after a period of time $t1$, which is a
predefined threshold, the node searches the data in the next zone and
this process continues until it gets the needed data or when it
reaches the maximum transmission range. If we are not able to find the
required data within the transmission radius $R$ the request is
directly given to the server. The time out interval set for each zone
is different to minimize the waiting time. 
\vskip 2mm
The power needed for each
transmission is assumed to be different. Initially, the transmission
power is kept at the minimum level and the node will search for the
data in the lowest transmission range. If we could not find the
desired data, the transmission power is increased to search for the
data in the next zone. Currently in our algorithm, power is increased
only by a fixed amount. The process of cache discovery is fully
distributed and runs in all the nodes in the network.
\subsection{Cache Management}
Cache management involves cache placement and replacement. When there
is a local miss the data item is fetched either from the neighboring
nodes or from the server.  The cache placement module is triggered
when the data item is brought in, to decide whether to cache or not
the incoming data. In order to cache more distinct data the caching
decision is done based on two parameters, size and distance. We set a
threshold value $T$, which is 50 \% of the cache size for a data item
to be admitted to cache. The data coming from the neighboring nodes
are also not cached in order to increase the data accessibility.
\vskip 2mm
In cooperative caching if data replacement decision is made by
individual nodes by considering only their local cache, the
performance is degraded because the data may be present in the
neighboring nodes. In order to cache more distinct data, new data item
fetched from adjacent nodes are not cached. When the cache is full,
appropriate data from the cache have to be evicted to make room for
the incoming data. The replacement policy proposed here considers the
number of references for a particular data item and gives more
emphasis data items that are referenced more than once. If we have
data items referenced only once then that set is given priority for
replacement. For this LRU policy is used. If an item is referenced
more than once the inter arrival time between the recent two
references is considered for eviction.
Let $t_c$ be the current reference time and $t_r$ be the previous reference time then $T_{{\rm int}}$, the inter arrival time is given by (\ref{eq2})
\begin{equation}\label{eq2}
T_{{\rm int}} = t_c- t_r
\end{equation}
If $t_c- t_r  = 0$, an  item whose last reference time  is smaller is
replaced.
If $T_{{\rm int}} > 0$, then the replacement decision is made on the
value of $K (i)$ which is given as (\ref{eq3})
\begin{equation}\label{eq3}
K(i) = {\rm Max} \sum_{i=1}^n t_c- t_r
\end{equation}
The data items with maximum inter arrival time is considered for
replacement. In both cases if more than one data item have the same
value, the TTL parameter is taken and the one with lower TTL value is
removed as the data with lower TTL will be outdated soon.
\section{\uppercase{Implementation}}
We have developed a simulation model in JAVA. The proposed scheme is a
fully distributed scheme, where each node runs an application to
request, retrieve and cache data items from other neighboring
nodes. Each node caches part of the requested items temporarily. The
simulated mobile environment consists of a number of mobile nodes
which are randomly placed in an area of 1000 $\times$ 1000 m$^2$. Each
node is identified by a node id and a host name. The position of each
node is given by the $x$ and $y$ coordinates. The data centre is
implemented in a fixed position in the simulation area.  The data
center contains all the data items requested by the mobile nodes. The
size of each data item is uniformly distributed between
$s_{{\rm min}}$ and $s_{{\rm max}}$.
\vskip 2mm
 The database in the data center
contains 1000 data items, with each item identified using $a$ data
id. The nodes that generate data request are selected randomly and
uniformly. The time interval between  two consecutive  queries
generated  from  each  node  follows  an  exponential distribution
with  mean  of 10 sec. Each   mobile node generates a single stream of
read only queries. The queries generated follows a Zipf distribution
\cite{r14}, which is frequently used to model non uniform
distribution. The data request is processed in FCFS manner at the
server. An infinite queue is used to buffer the request when the data
center is busy. Each miss in the cooperate cache will incur a delay of
8 ms to retrieve data from the data center.
\vskip 2mm 
Initially, the mobile nodes are randomly distributed in the simulation
area. After that each node randomly chooses its destination with a
speed $s$ which is uniformly distributed $U$
$(V_{{\rm min}},V_{{\rm max}})$ and travels with that constant speed
$s$. When the node reaches destination, it pause for 200
seconds. After that it moves to the new destination with speed
$s'$. The details of the simulation parameters are given in table
\ref{tab1}. 
\subsection{Simulation Parameters} 
\begin{table}[ht!]
\centering
\caption{Simulation Parameters}\label{tab1}
{\small
\begin{tabular}{|l|l|}
\hline
Parameter & Value\\ \hline
Simulation Time & 3600 sec \\ \hline
Simulation Area & 1000 $\times$ 1000 m$^2$\\ \hline
Database  & 1000 items\\ \hline
Cache size & 20--70 \% of total \\ \hline
           & database size\\ \hline
$S_{{\rm min}}$ & 1\\ \hline
$S_{{\rm max}}$ & 10\\ \hline
Nodes Density & 20--70\\ \hline
Mobility model & Random waypoint\\ \hline
Transmission Range & 500 m\\ \hline
Speed of the mobile host & 1--10 \\ \hline
Pause time & 200 sec\\ \hline
Mean query generation time & 10s\\ \hline
\end{tabular}}
\end{table}
\subsection{Metrics}
The performance metrics evaluated includes cache hit ratio, message
overhead and power savings ratio. The evaluation of these parameters
are done by varying the number of cache locations  with respect  to
number of nodes and the behavior of cache hit ratio for different
cache sizes. The hit ratio is defined as the percentage of requests
that can be served from previously cached data. Since the replacement
algorithm decides whether to cache the data or not, it affects the
cache hits of future requests. The percentage of power saved is
calculated as power usage of cooperative caching scheme compared-power
usage of the proposed scheme/power usage of the proposed
scheme. Message overhead is the overhead messages needed to manage the
cache discovery process in cooperative cache.
\section{\uppercase{Performance Evaluation}}
\begin{figure*}[ht!]
\centering
\epsfig{file=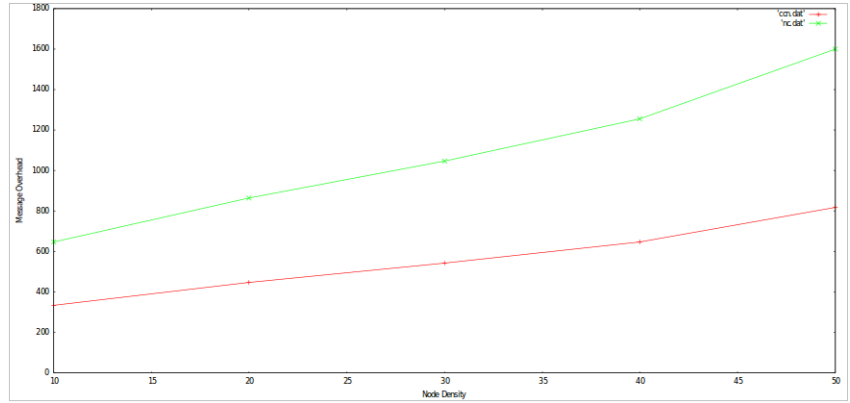,width=.70\linewidth} 
\caption{Message Overhead for Different Node Densities}\label{f1}
\end{figure*}

\begin{figure*}[ht!]
\centering
\epsfig{file=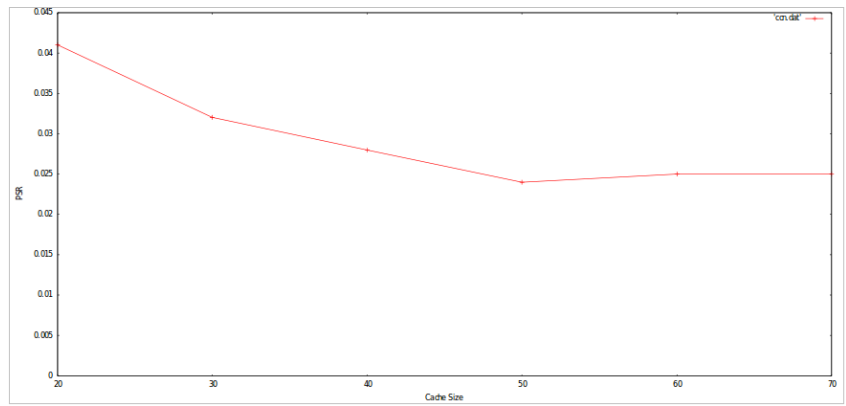,width=.70\linewidth} 
\caption{Power Savings Ratio for Different Node Densities}\label{f2}
\end{figure*}

\begin{figure*}[ht!]
\centering
\epsfig{file=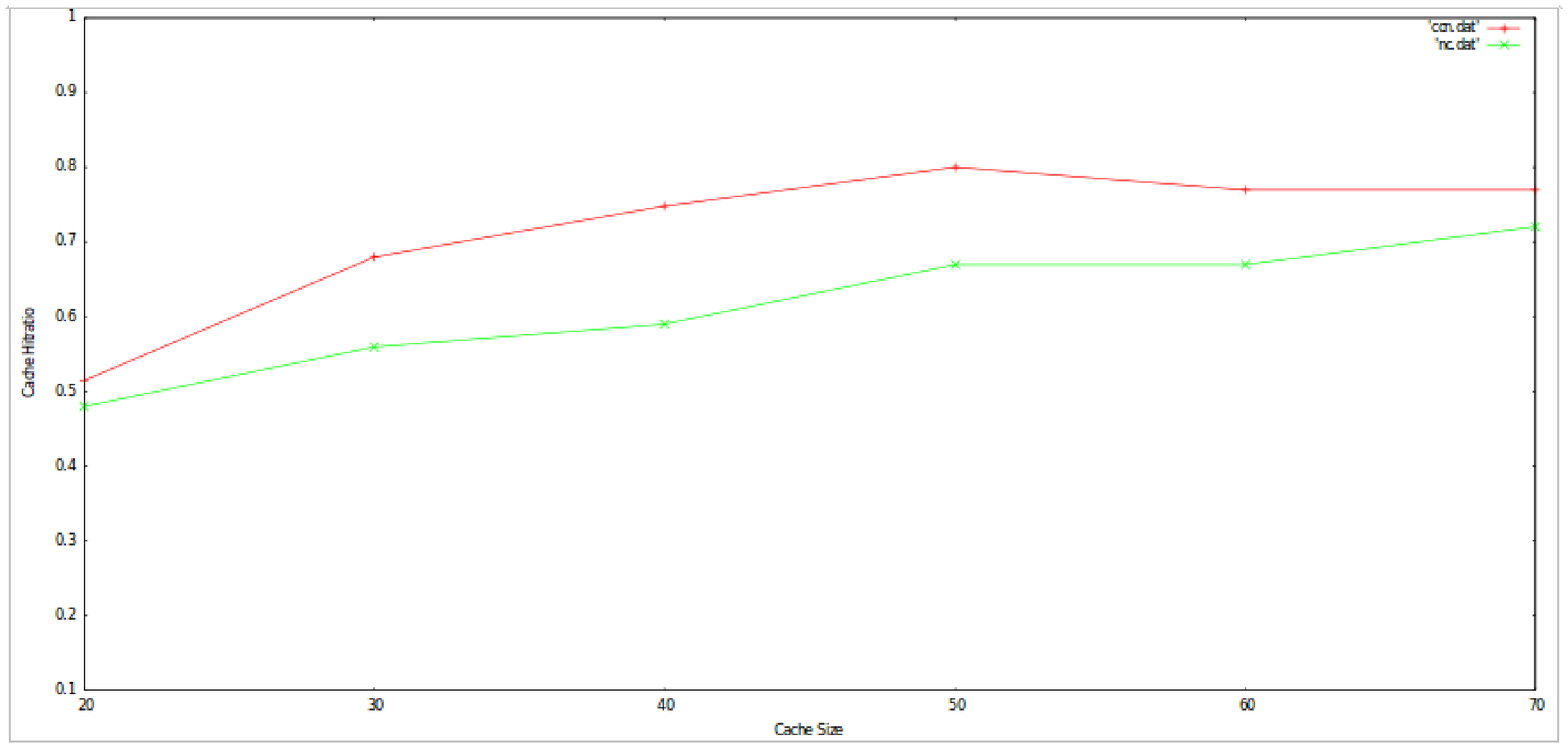,width=.70\linewidth} 
\caption{Cache Hit Ratio for Different Cache Sizes}\label{f3}
\end{figure*}

To evaluate the performance of the proposed cache discovery scheme, we
compared the performance of our new cooperative caching scheme
(CCN), with Neighbor Caching (NC), a caching scheme which uses
broadcasting and LRU for cache replacement. Figure \ref{f1} shows the
performance comparison of two schemes, as a function of message overhead
for different node densities. The figure shows that CCN outperforms NC
at all node densities. As the node density increases, the difference
become more significant, this implies that CCN can benefit from larger
node densities. Figure \ref{f2} depicts the power savings ratio of the
proposed cache discovery scheme compared to neighbor caching. From Figure
\ref{f3} we can see that the cache hit ratio for CCN is greater than
NC for different cache sizes. The relative performance of cache hit
ratio remains relatively stable for higher cache sizes.
\section{\uppercase{Conclusions}}
 In this paper, a cache discovery algorithm for cooperative caching
 for optimizing the power usage and decreasing energy consumption by
 reducing caching overhead and transmission range adjustment is
 proposed. The proposed algorithm is compared with other cooperative
 caching protocol in terms of message overhead and cache hit
 ratio. The power savings ratio of the proposed approach is also
 calculated.  The objective of our problem was to reduce the message
 overhead and the power consumption for data discovery. We designed a
 data discovery process based on the position of the neighboring nodes
 and according to the position of neighboring nodes the transmission
 range is adjusted for data retrieval. 
\small\balance

\noindent{\includegraphics[width=1in,height=2in,clip,keepaspectratio]{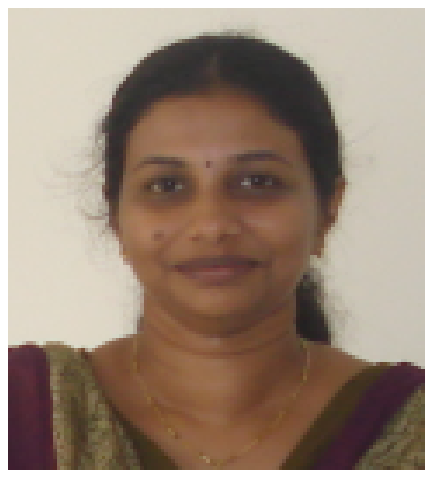}}
\begin{minipage}[b][1in][c]{1.8in}
{\centering{\bf {Preetha Theresa Joy}} is a Research Scholar in the Department of Computer Science at Cochin University of Science and Technology, Cochin, Kerala State, India. She received her M.Tech in Computer Science from Cochin University of  Scie-}\\
\end{minipage} Science and Technology. Her research interests include Computer Networks, Mobile Computing and Mobile Ad hoc Networks.\\\\\\
\noindent{\includegraphics[width=1in,height=1.7in,clip,keepaspectratio]{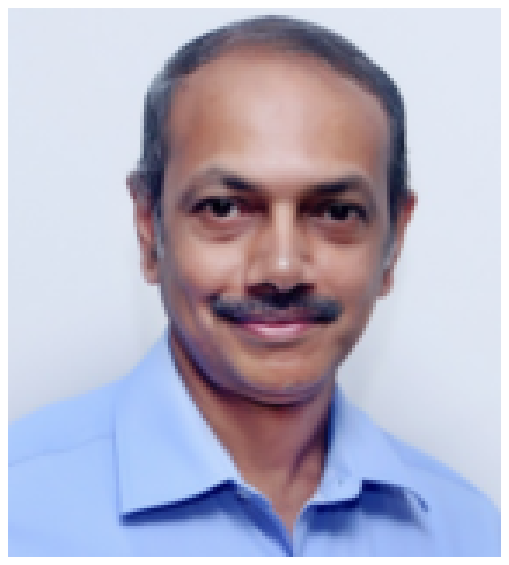}}
\begin{minipage}[b][1in][c]{1.8in}
{\centering{\bf{Dr. K Poulose Jacob, }}Professor of Computer Science at Cochin University of Science and Technology (CUSAT) since 1994, is currently the Pro Vice Chancellor. He has presented research papers in several International Conferences in Europe,   } \\
\end{minipage} USA, UK,   Australia and other countries. He has delivered invited talks at several national and international events. Dr. Jacob is a Professional member of the ACM (Association for Computing Machinery) and a Life Member of the Computer Society of India. He has more than 90 research publications to his credit. His research interests are in Information Systems Engineering, Intelligent Architectures and Computer Networks. \\\\
\small
\balance
\end{document}